**Measurement of Dosimetrical Cross Sections with 14.05 MeV Neutrons From Compact Neutron Generator**

Michal Kostal[1], Tomáš Czakoj[1], Jan Šimon[1], Marek Zmeškal[1], Martin Schulc[1], Alena Krechlerová[1], Tomáš Peltán[1], Filip Mravec[2], František Cvachovec[2], Vojtěch Rypar[1], Radim Uhlář[3], Petr Alexa[3], Ondřej Harkut[3], Zdeněk Matěj[2]

[1] Research Center Rez, Husinec-Rez, 250 68, 130, Czech Republic

[2] Masaryk University, Botanická 15, Brno 612 00, Czech Republic

[3] Department of Physics, VŠB-Technical University of Ostrava, 17. listopadu 2172/15, 708 00, Ostrava, Czech Republic

Email: Michal.Kostal@cvrez.cz

Telephone: +420266172655

**Abstract**

Dosimetry cross sections are fundamental quantities necessary for neutron dosimetry using the neutron activation method. It is worth noting that the uncertainty in cross sections is the major source of uncertainty in calculational predictions using nuclear data in simulations. Thus, cross section validation is a key issue in any aims for refinement of any predictions. A small compact neutron generator is a promising tool for performing integral experiments and even for differential experiments. This paper deals with the measurement of the differential dosimetry cross sections using a small compact D-T neutron generator with 14.05 ± 0.08 MeV neutron emission ($1\times10^8$ n/s into 4π). Achieving measurable activation at such a low flux field is allowed by using a larger amount of activation material placed in close measurement geometry during decay gamma measurement. The experimentally determined cross sections are in good agreement with the cross sections in the IRDFF-II dosimetry library. The comparison with other nuclear data libraries was performed as well. Its worth noting, the mean standard deviation in IRDFF-II library is about 4 %, while in case of other data libraries they are from 5.5 % – 7.5 %. This result can be understood as a validation of IRDFF-II using 14.05 MeV neutrons and also a confirmation of the applicability of small compact generators in the measurement of activation cross sections.

# 1 Introduction

Neutrons produced by the $^{3}$H($^{2}$H,n)$^{4}$He reaction are ideal for the measurement of cross-sections because if the neutrons are produced in a fixed-target generator, their spectrum has the character of a narrow peak with well-defined energy [1]. For the purpose of measurement of dosimetry cross sections, generators with a high achievable flux in the irradiated target are often used. High flux can be reached either by a close position to the target or by high neutron emission. Namely, there is reported use of generators with emission of the order of $3\times10^{10}$ n/s [2], [3] to $2.5\times10^{11}$ n/s [4] or generators with smaller emission but close position to the target where the flux in measuring position is of order $1\times10^7$ cm$^{-2}\cdot$s$^{-1}$ [5] to $1\times10^8$ cm$^{-2}\cdot$s$^{-1}$ [6].

The small D-T generators are useful and versatile tools used in various fields, such as science or industry. There are reported many applications using active measurements like planetary science [7] or geology [8],[9], homeland security [10], industry [11], [12], or for calibration of detectors used in fusion research [13].

The present paper deals with the measurement of cross sections with a low emission generator ~ $1\times10^8$ n/s with the highest achievable neutron flux at the foil position of about $3\times10^5$ $cm^{-2}\cdot s^{-1}$. The foils were placed on the tube surface at a distance of 5.25 cm from the tritium/deuterium target irradiated with deuterons/tritons emitted from the Penning source [14]. The low activation of the foil is compensated by close measuring geometry of the irradiated samples, which allows measurement of even low active samples. A precise activity measurement is possible using a validated mathematical model of the HPGe detector which enables the determination of the efficiency curve by calculation. Thanks to that approach, it is possible to measure gamma activities of any sample in arbitrary measuring geometry.

## 2 Experimental setup

The experiments were carried out using a low neutron emission MP-320 Neutron Generator [15] (NG) produced by Thermo Fisher Scientific Inc and operated by VŠB - Technical University of Ostrava. It is a portable NG (sealed tube type) with a cylindrical shape (diameter of 12 cm in reinforced parts, 10.5 cm in others, and length of 57 cm). Declared neutron emission is $1\times10^8$ n/s into $4\pi$. The NG used is located in an underground laboratory with dimensions of 375 cm × 240 cm, and a height of 270.5 cm. The NG target in this experiment was placed in the axis of the laboratory 106 cm above the floor, 120 cm from the side walls, and 109 cm from the back wall. The computational model of this generator, depicted in Figure 2, was compiled using a radiogram of the generator [16]. The irradiated foils were taped on the generator tube in the target plane, see Figure 1. Setting the foils in such positions ensures that the foils are positioned symmetrically at 90° regarding the deuteron/triton beam, at the same distance from its center. This selected position has the highest possible neutron flux because this position is the closest possible to the target.

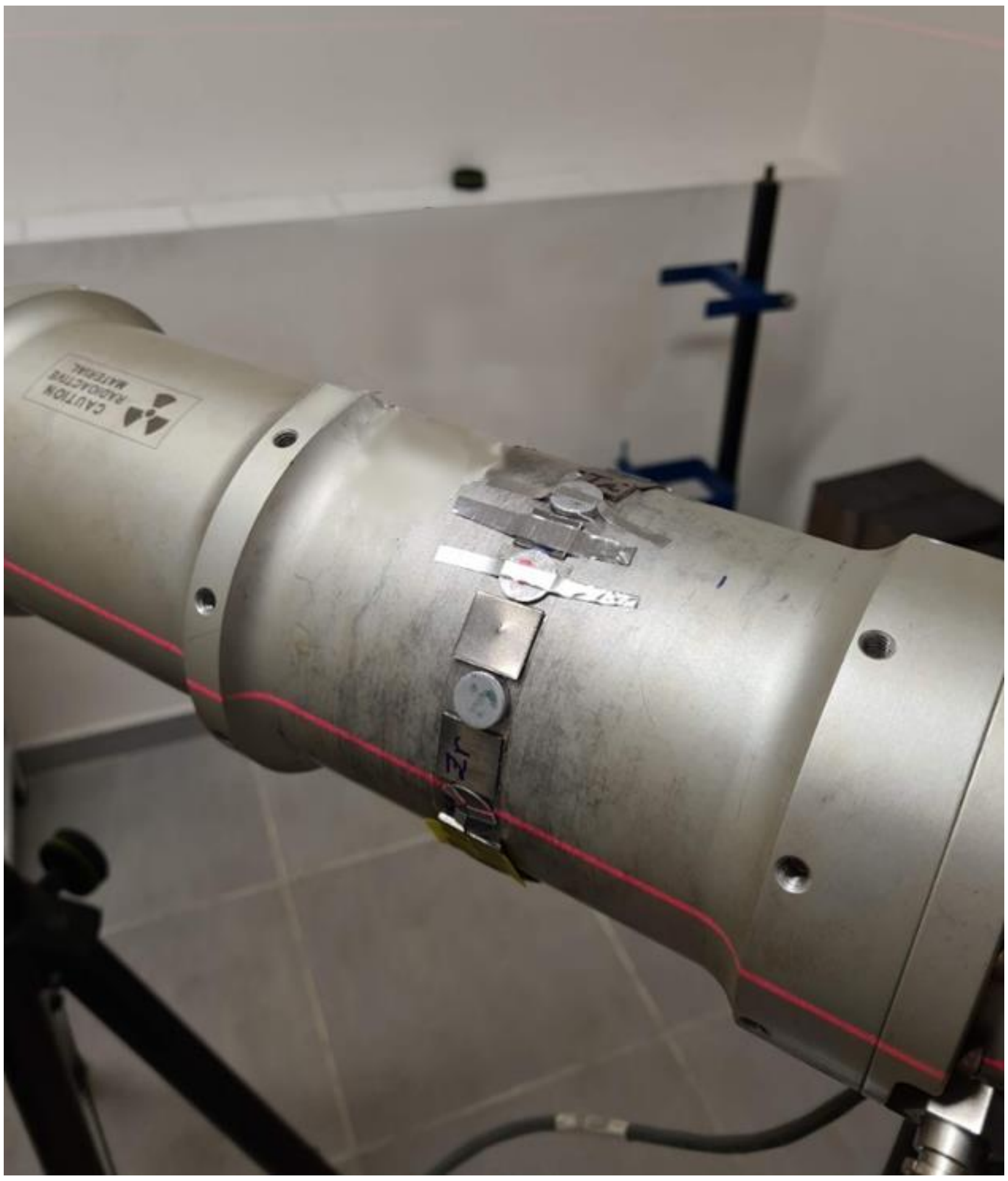

Figure 1: View on DT generator with foils

The neutron generator can be operated at 100 – 240 V, 50 – 60 Hz AC or 24 V DC at 6 A (during the experiments 230 V 50 Hz AC were used). The deuteron (triton) energy range is fixed by the operating acceleration voltage that spans the interval from 30 kV to 90 kV. In this experiment, the operating voltage was set to 80 kV, thus the deuteron (triton) energy at the interaction point with the triton (deuteron) spans to interval from 0 keV to 80 keV due to the deuteron (triton) stopping in the tritiated (deuterated) target [14].

The neutron spectrum was measured by stilbene scintillation spectrometry at the angle of 90° and by activation foils. Due to the presence of deuterium in the target, there are also D-D neutrons in the neutron field formed by the generator as well as T-T neutrons due to the presence of tritons in the accelerated beam. It should be noted that the d-d cross section is about 300 times smaller than the d-t cross section in the region of the operating voltage of 80 kV, and therefore, it represents only a minor effect. Similarly, the t-t cross section is about 400 times smaller [17].

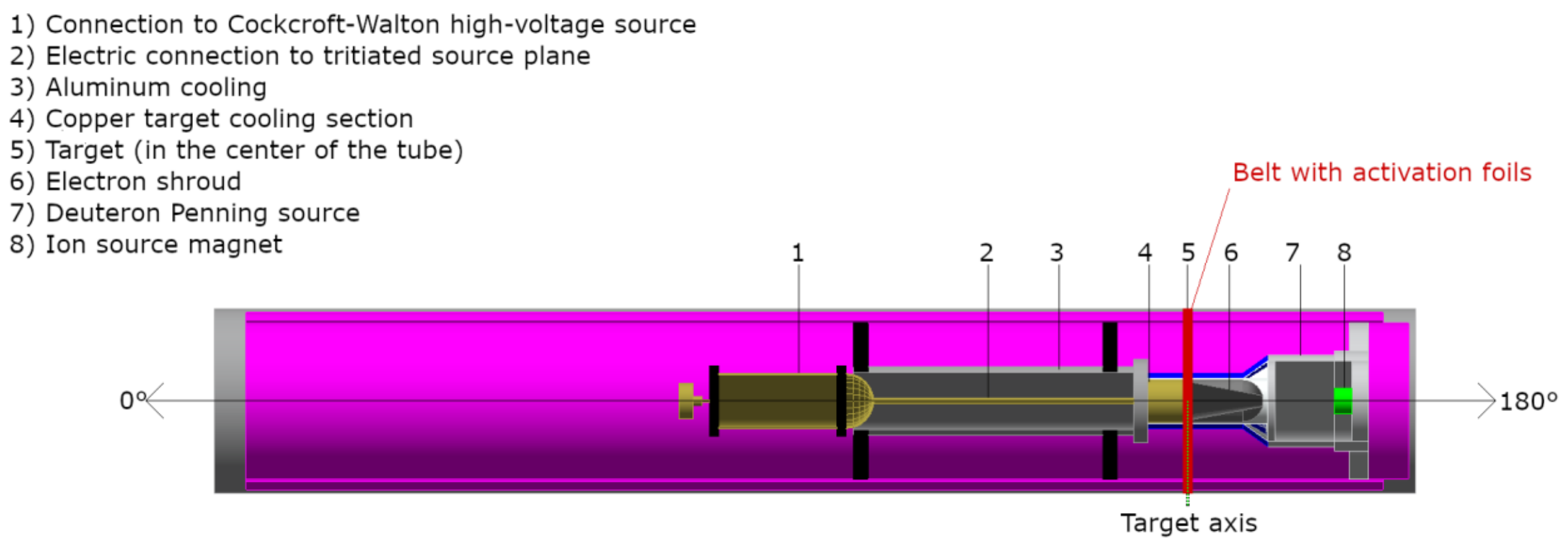

Figure 2: Schematic view of MP-320 Neutron Generator

## 3 Experimental and calculation methods

The shape of the neutron spectrum in the energy range of 1.1 MeV to 15.3 MeV was measured via the proton-recoil method using a stilbene scintillator with neutron and gamma pulse shape discrimination. These measurements were carried out during the irradiation of selected dosimeters. The mean neutron energy was determined to be $14.05 \pm 0.08$ MeV, which was derived by the fit of the experimental spectrum. The value is in agreement with relativistic calculations that predict slightly lower value of the neutron energy for the neutron emission angle of $90^{o}$ compared to the non-relativistic calculations. For accelerated deuterons interacting with tritons at rest one gets 14.06 MeV and 14.08 MeV, respectively, and for accelerated tritons interacting with deuterons at rest one gets 14.05 MeV and 14.06 MeV, respectively. Moreover, there is another effect connected to particle slowing down in the Zr target of the neutron generator. In the calculations we assume the target thickness of 2 μm, its diameter of 1 cm, density 6.57 g/cm$^3$ and an equal load of D and T. Deuteron and triton stopping powers are calculated using SRIM-2013.00 code [18]. Mean neutron energies and their standard deviations are obtained from dt-cross section data [19] and calculated using the method described in [20]. The obtained value $14.05 \pm 0.03$ MeV agrees with the experimentally determined neutron energy.

The activities induced in the foils during irradiation were determined using semiconductor HPGe spectrometry. Due to the low induced activity, a close geometry was used, with the foil on the HPGe cap. The non-constant irradiation schedules were taken into account using the known time profile of the neutron emission. The geometrical effect of different dimensions of the foils, which result in variations of the neutron flux between the foils, was solved by means of calculations.

### 3.1 Spectrum measurement

The neutron spectrum was characterized using the proton recoil method and stilbene organic scintillator. This kind of detector allows simultaneous measurement of neutrons and gammas because it is sensitive to both types of particles. Separation of both signals is performed by means of pulse shape discrimination (PSD). Its principle consists of comparing the area limited by the part of the trailing edge of the measured response and the area limited by the entire response. The mentioned areas are simply the time integrals of the measured response. The separation of the measured signal is depicted in Figure 3. It can be seen that gammas (left part, propagated by Compton scattered electrons) and neutrons (right part, propagated by recoiled protons) are well separated.

The obtained recoiled protons are then deconvoluted to neutron flux by the Maximum Likelihood Estimation [21]. The robustness of the used methodology was tested in various reactor fields as well as in accelerator fields [22].

A crystal with a diameter of 45 mm and a length of 45 mm was used in the measurement. Such crystal is usable due to a relatively low gamma background during irradiation.

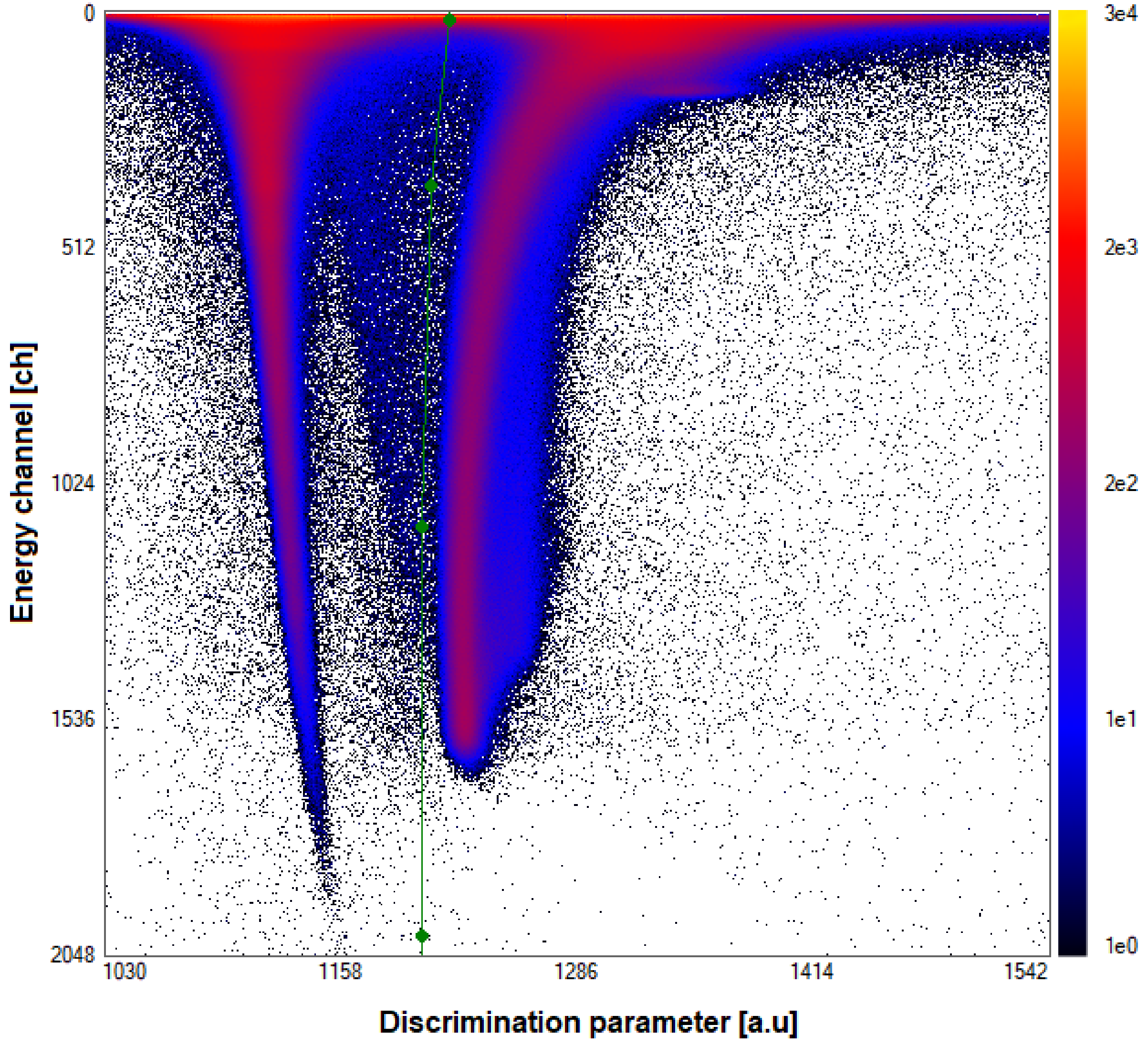

Figure 3: PSD of the two parametric spectra in the field of MP-320 neutron generator. The left peak is the gamma peak (lower discrimination parameter), and the right peak is the neutron peak, the green curve is the separation line.

## 3.2 HPGe spectrometry

HPGe spectrometry was used to determine the activities of specific nuclei that originated in foils during their irradiation. The measured quantity was the Net Peak Area (NPA) of the studied activation product. Due to the low flux in the irradiation position, the induced activity is in the order of 1 Bq in the case of $^{59}$Fe induced from $^{59}$Co(n,p) to about 200 Bq in the case of $^{196}$Au induced in $^{197}$Au(n,2n). The end cap measuring geometry was applied to obtain experimental values with low uncertainty (see Figure 4). This is enabled thanks to the well described HPGe detector with a precise mathematical model [23], allowing the use of computational determinations of the efficiency and Coincidence Summing Correction Factors (CSCF) [24]. Due to such approach, it is possible to use a dosimeter in any measuring geometry. The tabulated branching ratios and decay constants from IRDFF-II [25] were used. The calculated values of efficiency and CSCF for the dosimeters used are given in Table 1.

In the case of the In foil, the NPA was also measured in the well of an ultra-low background HPGe spectrometer GWD-3023 at the underground laboratory of VŠB – Technical University of Ostrava [26]. The determined In specific activity at the end of irradiation differed by 0.9 % what is highly below the uncertainty.

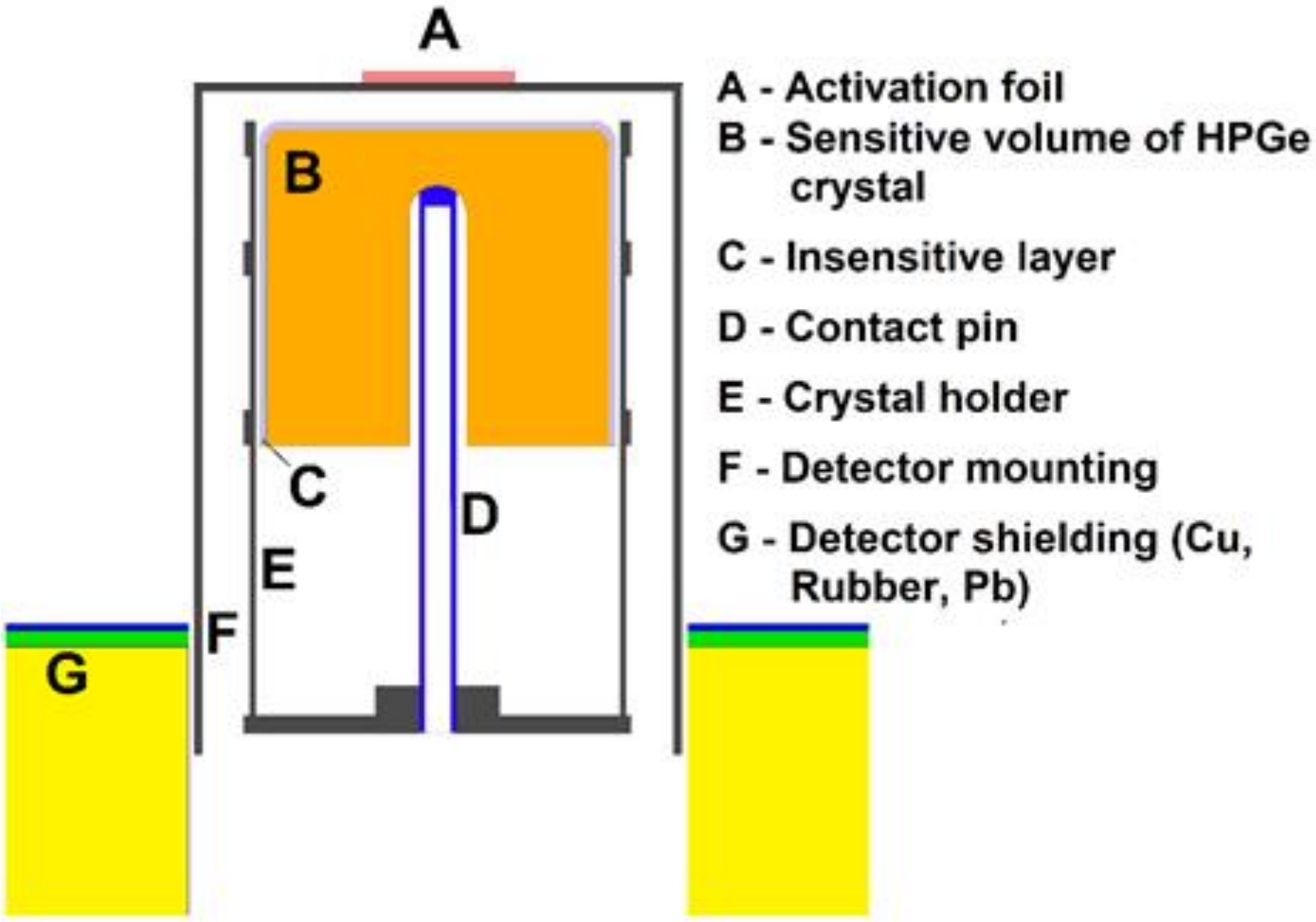

Figure 4: Scheme of measuring geometry

Table 1.: Efficiencies and CSCF for used dosimeters

| Reaction | Dimensions [cm] | Peak [keV] | Efficiency | $k_{CSCF}$ |
|---|---|---|---|---|
| $^{27}$Al(n,α) | D=1.27, th. 0.3 | 1368.6 | 2.899E-2 | 0.863 |
| $^{24}$Mg(n,p) | D=1.8, th. 0.5 | 1368.6 | 3.156E-2 | 0.843 |
| $^{90}$Zr(n,2n) | 1.3 × 1.3 × 0.2 | 909.15 | 4.062E-2 | 0.903 |
| $^{58}$Ni(n,x)$^{57}$Co | 1.6 × 1.6 × 0.3 | 122.0 | 1.084E-1 | 1.000 |
| $^{58}$Ni(n,p) | | 810.8 | 4.067E-2 | 0.936 |
| $^{47}$Ti(n,x)$^{47}$Ti | 2.1 × 2.8 × 0.127 | 159.4 | 1.606E-1 | 1.000 |
| $^{46}$Ti(n,p) | | 889.3 | 4.145E-2 | 0.826 |
| $^{48}$Ti(n,p) | | 983.5 | 3.828E-2 | 0.660 |
| | | 1037.5 | 3.664E-2 | 0.653 |
| $^{89}$Y(n,2n) | D=2.54, th. 0.127 | 898.0 | 4.147E-2 | 0.846 |
| | | 1836.1 | 2.287E-2 | 0.824 |
| $^{93}$Nb(n,2n) | D=1.8, th. 0.1 | 934.4 | 4.096E-2 | 1.000 |
| $^{115}$In(n,n') | D=1.27, th. 0.1 | 336.2 | 9.832E-2 | 1.000 |
| $^{197}$Au(n,2n) | D=1.5, th. 0.1 | 333.0 | 7.463E-2 | 0.844 |
| | | 355.0 | 7.305E-2 | 0.990 |
| $^{51}$V(n,α) | 1 × 1 × 0.06 | 983.5 | 4.124E-2 | 0.643 |
| | | 1037.5 | 3.968E-2 | 0.635 |
| $^{56}$Fe(n,p) | 1.3 × 1.1 × 0.1 | 846.8 | 4.504E-2 | 0.935 |
| | | 1810.7 | 2.403E-2 | 0.803 |
| $^{59}$Co(n,α) | D=1.5, th. 0.075 | 846.8 | 4.541E-2 | 0.934 |
| | | 1810.7 | 2.417E-2 | 0.802 |
| $^{59}$Co(n,p) | | 1099.2 | 3.685E-2 | 1.000 |
| | | 1291.6 | 3.231E-2 | 1.000 |
| $^{59}$Co(n,2n) | | 810.8 | 4.702E-2 | 0.933 |
| $^{55}$Mn(n,2n) | 1.4 × 0.85 × 0.1 | 834.8 | 4.820E-2 | 1.000 |
| $^{169}$Tm(n,2n) | 1.8 × 1.2 × 0.01 | 815.9 | 4.856E-2 | 0.629 |

Activities at the end of irradiation $A_{\text{End}}$. were determined from the experimental NPA using Eq. 1. The experimental reaction rate $q$ can be evaluated from the activity with knowledge of the irradiation scheme (see Sect. 3.3). The experimental cross section $\sigma$ is determined from the reaction rate with knowledge of the neutron flux $\phi$ in the target. The neutron flux was derived from the monitoring reaction $^{27}$Al(n,α). This reaction was used for its good knowledge and low uncertainties in its cross section [25].

$$A_{\text{End}} = NPA(T_{\text{Meas.}}) \times \frac{\lambda}{\varepsilon \times \eta} \times \frac{1}{(1 - e^{-\lambda . T_{\text{Meas.}}})} \times \frac{1}{e^{-\lambda . \Delta T}} \times \frac{1}{k_{\text{CSCF}}} \qquad (1)$$

$$q(\overline{P}) = A_{\text{End}}/N \times \left(\frac{A(\overline{P})}{A_{\text{Sat}}(\overline{P})}\right)^{-1} \quad (2)$$

$$\frac{A(\overline{P})}{A_{\text{Sat}}(\overline{P})} = \sum_i Q_{\text{rel}}^i \times \left(1 - e^{-\lambda.T_{\text{Irr}}^i}\right) \times e^{-\lambda.T_{\text{end}}^i} \quad (3)$$

$$\sigma(E) = \frac{q(E)}{\varphi(E)} \quad (4)$$

where:

$q$; is the reaction rate of activation during irradiation batch;

$\varphi$; is the neutron flux in the target;

$\sigma$; is the experimental cross section;

$\lambda$; is the decay constant of the radioisotope considered;

$T_{\text{Meas.}}$; is a time of measurement by the HPGe;

$Q_{\text{Rel.}}(i)$; relative emission of neutron generator with current $i$

$\Delta T$; is the time between the end of irradiation and the start of the HPGe measurement;

$NPA(T_{\text{Meas.}})$; is the Net Peak Area (the measured number of counts);

$\varepsilon$; is the gamma branching ratio;

$\eta$; is the detector efficiency (it's being determined via MCNP6.2 calculation);

$N$; is the number of target isotope nuclei;

$k_{\text{CSCF}}$ is coincidence summing correction factor;

$T_{\text{Irr.}}$; are times of individual irradiation periods;

$T_{\text{end}}$; are times between end of individual irradiation periods and end of irradiation.

### 3.3 Neutron flux monitoring

At the end of irradiation, the neutron emission profile has a significant effect on the activity of resulting nuclei. During irradiation, a stable mode of 80 kV was used, when it can be assumed that the neutron emission is proportional to the deuteron (triton) current on the tritiated (deuterated) target. The time evolution of the target current, being proportional to the neutron emission profile used in Eq. (3), is shown in Figure 5.

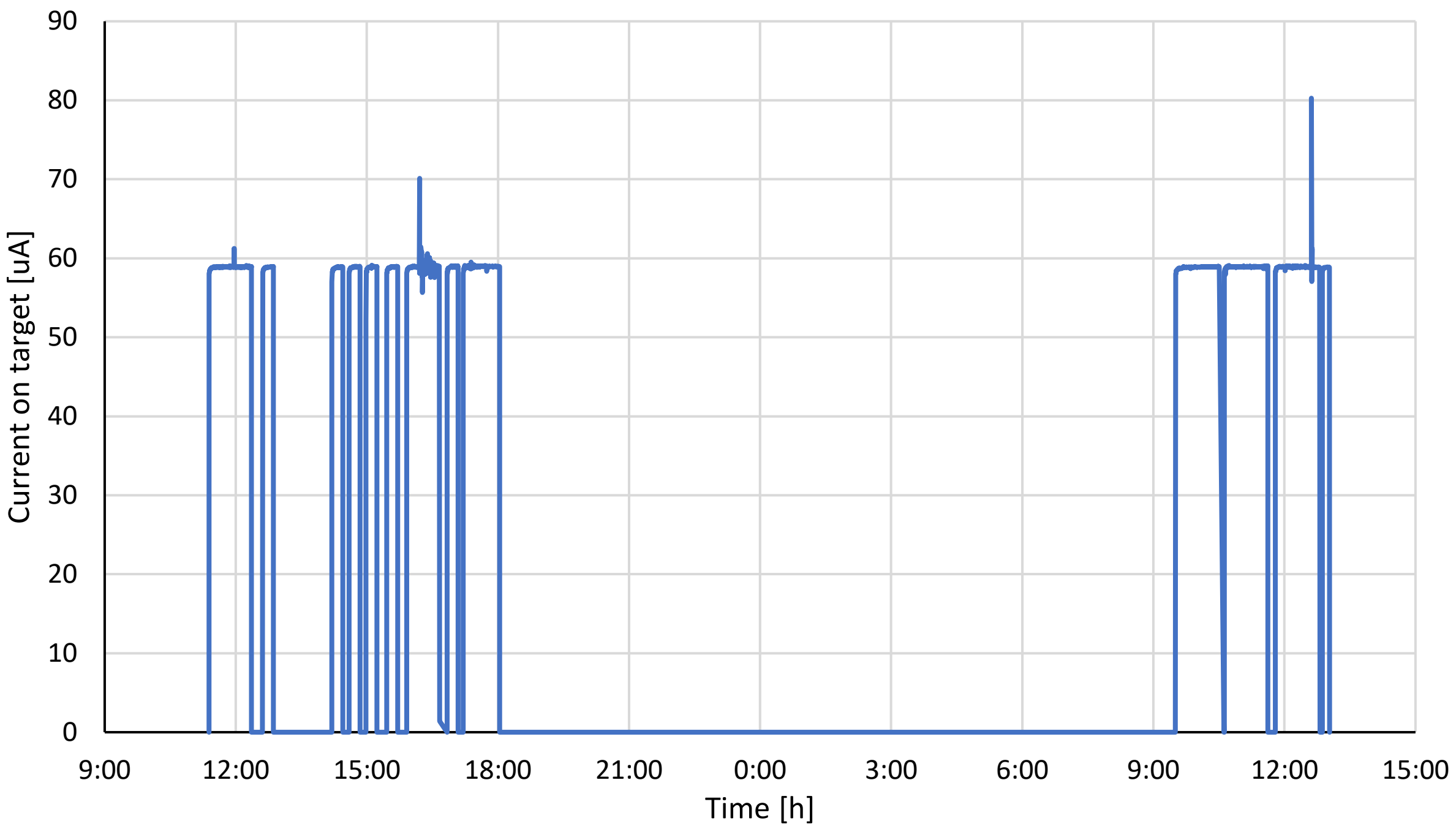

Figure 5: Neutron emission profile during the activation experiment

### 3.4 Calculations

The all-purpose Monte Carlo code Geant4 [27] in its version Geant4.11.0 was used to simulate the neutron flux in different activation detectors. A detailed model of the neutron generator was prepared using a X-ray image [16] and operational parameters. Its scheme is shown in Figure 2, in which the main construction parts are labeled. Neutrons are simulated by the direct reaction of deuterons (tritons) in a thin target made of zirconium tritide (zirconium deuteride). Nuclear data for deuterons were taken from the G4TENDL1.4 charged particle default library, where they are taken from ENDF/B-VIII.0 [17]. Data for neutron transport were also taken from this library prepared for Geant4 [28], [29], and [30]. This data was implemented through the QGSP_BIC_AllHP physics reference list. The yield of secondary neutrons is biased by the multiplication of the deuteron cross section through G4BOptnChangeCrossSection class.

## 4 Results

### 4.1 Spectrum

The neutron spectrum was characterized using the proton recoil method and organic scintillator and is shown in Figure 6. There are three major peaks in the spectrum, namely at 1.85 MeV, 2.85 MeV, and 14.05 MeV originating from D-D and D-T reactions. The uncertainties in the measured neutron spectrum are about 10% for each group.

The measured gamma spectrum is plotted in Figure 7. The related uncertainties are about 10% in peaks and about 20% in the minima between peaks.

Measurement of the neutron spectrum with a stilbene detector confirms the peak character of the neutron spectrum around the neutron generator, as can be seen in Figure 6. This means that in the case of reactions with a threshold above the first two peaks, the cross section in the D-T neutron peak, being 14.05 MeV, can be evaluated as there is no contribution from neutrons in 1.85 MeV and 2.85 MeV peaks. In the case of reactions with a threshold below 1.9 MeV, the activation product is induced by all 3 peaks, which are present in the neutron spectrum.

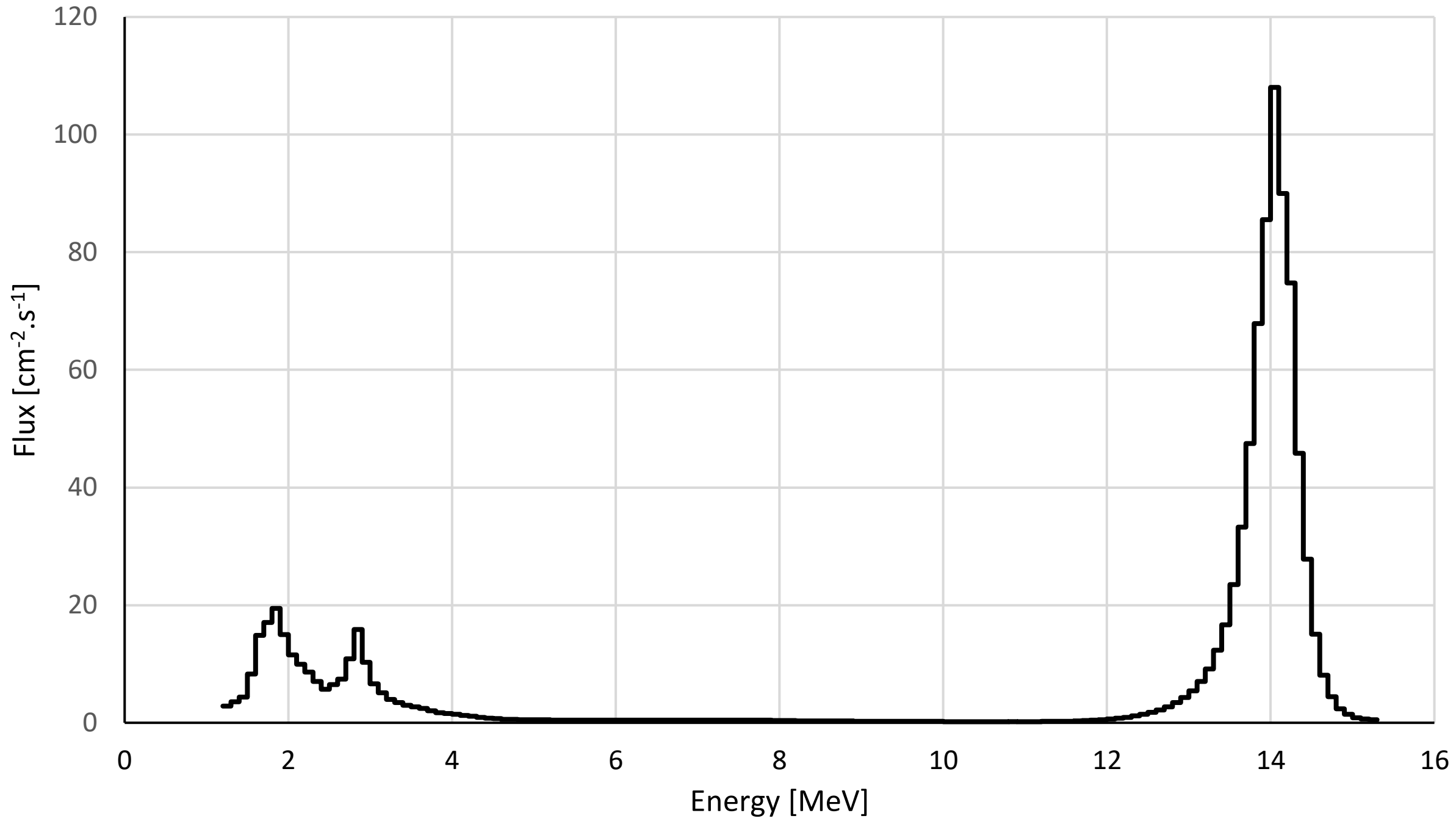

Figure 6: Neutron spectrum in target plane at the distance of 1 m from target center

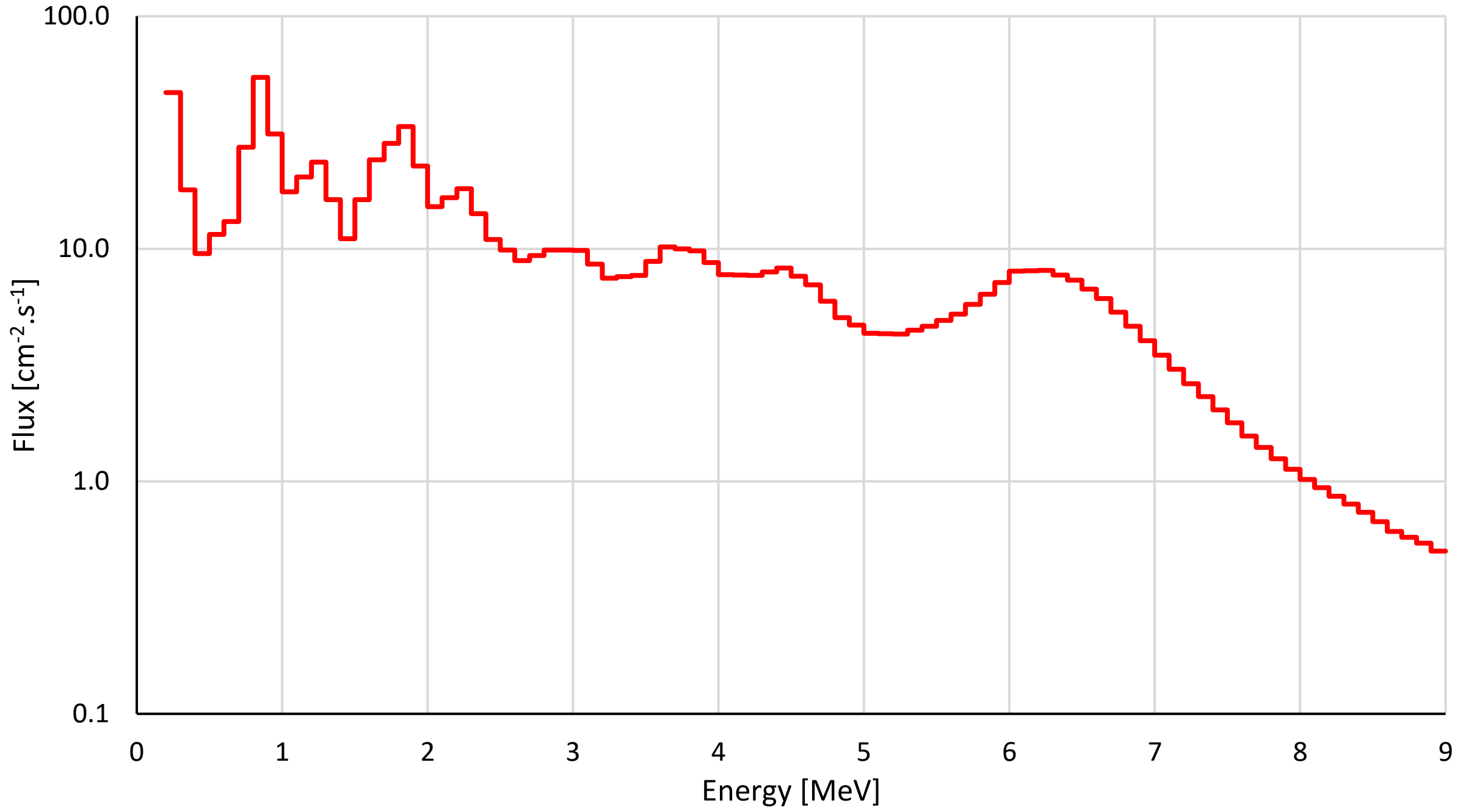

Figure 7: Measured gamma spectrum 1 m from the generator in the position of neutron spectrum measurement

The gamma to neutron ratio in regions above 1 MeV was experimentally determined to be ~ 0.7. Such a low ratio is accompanied by a low presence of high energy photons, which reflect a low flux of thermal neutrons. The high energy neutron field manifests itself in the gamma spectrum, causing the hydrogen capture peak to be relatively small, but the gamma peaks from inelastic scattering on silicon are significant, and the oxygen inelastic scattering peak at 6.2 MeV is also well visible. Due to the relatively low share of thermal neutrons (demonstrated by a relatively small hydrogen peak at 2.22 MeV compared to thermal spectrum like at thermal neutron driven reactors [31]), there is a small share of very high energy gammas (up to 8 MeV), thus (γ,n) reactions can be neglected.

### 4.2 Reaction rates

The reaction rates of each dosimeter placed on the surface of the neutron generator are given in Table 2. The related uncertainties include the NPA measurement uncertainty, position uncertainty on target, as well as the systematic uncertainty of the calculated efficiency being 1.9% (see Fig. 9 in [32]). The effect of (γ,n) reactions which leads to the additional production of the same residue as (n,2n) reactions, can be neglected due to the low high-energy gamma flux.

Table 2.: Reaction rates on D-T generator at 80 kV and current 59 μA

| Reaction | Reaction rate [$s^{-1}$] | Rel. Unc. [%] |
|---|---|---|
| $^{27}$Al(n,α) | 3.506E-20 | 2.6 |
| $^{nat.}$Ti(n,x)$^{47}$Ti | 4.239E-21 | 2.7 |
| $^{58}$Ni(n,p) | 1.112E-19 | 2.2 |
| $^{48}$Ti(n,p) | 1.845E-20 | 2.4 |
| $^{92}$Nb(n,2n)$^{92}$Nb$^{*}$ | 1.389E-19 | 2.7 |
| $^{89}$Y(n,2n) | 2.303E-19 | 2.8 |
| $^{115}$In(n,n') | 2.984E-20 | 3.4 |
| $^{56}$Fe(n,p) | 3.350E-20 | 3.0 |
| $^{197}$Au(n,2n) | 5.971E-19 | 3.2 |
| $^{24}$Mg(n,p) | 5.487E-20 | 3.4 |
| $^{nat.}$Ni(n,x)$^{57}$Co | 1.757E-19 | 2.2 |
| $^{90}$Zr(n,2n) | 1.970E-19 | 2.5 |
| $^{59}$Co(n,α) | 8.939E-21 | 3.8 |
| $^{59}$Co(n,2n) | 2.003E-19 | 2.3 |
| $^{59}$Co(n,p) | 1.570E-20 | 5.1 |
| $^{51}$V(n,α) | 4.493E-21 | 3.8 |
| $^{55}$Mn(n,2n) | 2.043E-19 | 2.7 |
| $^{169}$Tm(n,2n) | 6.327E-19 | 2.6 |
| $^{46}$Ti(n,p) | 8.069E-20 | 5.0 |

An interesting issue is the determination of neutron flux in lower energy peaks and the possibility of their measurement with activation detectors. The $^{115}$In(n,n’) reaction seems to be a good candidate for such measurement because of the significantly smaller cross sections in the 14.05 MeV region than in the 2 MeV region. Namely, the cross section averaged over the lower peaks is about 4.3 times larger than the cross section in the 14.05 MeV peak. This is reflected in the fact that the $^{115}$In$^{*}$, formed by the lower energy peaks, is giving 36 % of the total $^{115}$In$^{*}$. The neutron flux in lower energy peaks, evaluated from the stilbene measurement and correction of the flux calculation, gives $3.9 \cdot 10^4$ $cm^{-2}.s^{-1}$ (see Table 3) and it is in excellent agreement with the neutron flux derived from the In foil being $3.8 \cdot 10^4$ $cm^{-2}.s^{-1}$.

This flux can be used also to determine the lower peak contribution to the measured reaction rates, thus allowing to evaluate reaction rates of the lower energy threshold reactions as cross sections in the 14.05 MeV peak. This correction was determined to be 2.8% in the case $^{58}$Ni(n,p)$^{58}$Co and 1.2% for $^{nat.}$Ti(n,x)$^{47}$Sc, in the case of $^{46}$Ti(n,p) and $^{59}$Co(n,p), it can be neglected because activation from lower peaks is not higher than 0.02%.

The cross sections used in all evaluations and geometrical corrections were derived from the IRDFF-II data library and the ratio between peaks used in the weighting of spectrum average cross section was taken from stilbene measurement. The activation rate in the 14.05 MeV peak uses IRDFF-II cross sections and flux from the $^{27}$Al(n,α) monitor.

Table 3.: Neutron fluxes in lower energy peaks evaluated from stilbene measurement

| | Flux [$cm^{-2}.s^{-1}$] | |
|---|---|---|
| | Measured by stilbene | Evaluated in target |
| 1.8 - 1.9 MeV | 6.628E+1 ± 9 % | 2.651E+4 ± 9 % |
| 2.8 - 2.9 MeV | 3.104E+1 ± 11 % | 1.242E+4 ± 11 % |
| Total flux from foils | | 3.794E+4 ± 8 % |

As the foils have different thicknesses, the calculational correction to the actual flux was applied using Geant4. The prediction is in good agreement with the $1/r^2$ law. The neutron flux in the target was derived from the $^{27}$Al(n,α) reaction. The activation cross sections derived using the flux correction factor, are given in Table 4. Reaction rates of $^{58}$Ni(n,p) and $^{nat.}$Ti(n,x)$^{47}$Sc reaction, given in Table 4, were corrected to the contribution of lower energy neutrons as well. The presented uncertainty takes into account uncertainties in the actual foil measurement, monitor measurement, and monitor cross section.

As mentioned previously, the neutron flux at a distance of 1 m was measured during activation. It is worth noting that the value of neutron flux in the 14.05 MeV peak from the stilbene measurement, recalculated to the generator tube surface position, is in good agreement with the flux derived from the $^{27}$Al(n,α) monitor.

The evaluated experimental cross sections were compared with the data in the nuclear data libraries (see Table 5). It is worth noting that in IRDFF-II the agreement is good. For most of the evaluated cross sections, the difference between the experimental and tabulated data is not higher than related uncertainty.

In the case of $^{55}$Mn(n,2n), $^{89}$Y(n,2n), $^{90}$Zr(n,2n), and $^{197}$Au(n,2n), differences higher than one standard deviation were observed, thus they were compared with cross section measured by other authors in same energy region. In the case of $^{55}$Mn(n,2n), there were reported values of 3.6% lower [33], 7.0% higher [34], and 4.5% higher [35]. For $^{89}$Y(n,2n) reaction, there were evaluated higher results, namely 4.5% [36], 3.1% [37], and 0.9% [38]. For $^{90}$Zr(n,2n) reaction, there were reported values of 3.3% lower [39], 5.3% lower [40], and 0.8% higher [41]. In the case of $^{197}$Au(n,2n), there were reported values 5.5% higher [42], 3.7% higher [43], and 1.4% higher [44].

Table 4.: Evaluated cross section of selected reactions in 14.05 MeV

| | Flux correction factor | Evaluated XS [b] | Rel. Unc. [%] |
|---|---|---|---|
| $^{nat.}Ti(n,x)^{47}Sc$ | 1.019 | 0.0143 | 4.0 |
| $^{58}Ni(n,p)$ | 1.034 | 0.363 | 3.6 |
| $^{48}Ti(n,p)$ | 1.034 | 0.0619 | 3.7 |
| $^{93}Nb(n,2n)^{92}Nb^{*}$ | 1.034 | 0.466 | 4.0 |
| $^{89}Y(n,2n)$ | 1.005 | 0.795 | 4.0 |
| $^{56}Fe(n,p)$ | 1.038 | 0.112 | 4.1 |
| $^{197}Au(n,2n)$ | 1.034 | 2.004 | 4.3 |
| $^{24}Mg(n,p)$ | 1.027 | 0.185 | 4.4 |
| $^{58}Ni(n,x)^{57}Co$ | 1.038 | 0.587 | 3.6 |
| $^{90}Zr(n,2n)$ | 1.038 | 0.658 | 3.8 |
| $^{59}Co(n,\alpha)$ | 1.034 | 0.0300 | 4.8 |
| $^{59}Co(n,2n)$ | 1.034 | 0.672 | 3.7 |
| $^{59}Co(n,p)$ | 1.034 | 0.0527 | 5.8 |
| $^{51}V(n,\alpha)$ | 1.038 | 0.0150 | 4.8 |
| $^{55}Mn(n,2n)$ | 1.034 | 0.686 | 4.0 |
| $^{169}Tm(n,2n)$ | 1.045 | 2.100 | 3.9 |
| $^{46}Ti(n,p)$ | 1.034 | 0.271 | 5.8 |

Table 5.: Comparison between measured values, and library value by means Eval./E - 1 [%]

| | IRDFF-II | ENDF/B-VIII.0 | JEFF-3.3. | JENDL-4 | CENDL-3.2. |
|---|---|---|---|---|---|
| $^{nat.}Ti(n,x)^{47}Sc$ | 2.3 | -11.0 | -7.3 | 8.5 | -19.9 |
| $^{58}Ni(n,p)$ | -1.2 | -4.8 | 5.4 | -11.5 | 3.2 |
| $^{48}Ti(n,p)$ | 0.8 | -4.4 | 0.7 | -2.4 | -1.7 |
| $^{93}Nb(n,2n)^{92}Nb^{*}$ | -1.4 | - | - | - | - |
| $^{89}Y(n,2n)$ | 6.2 | 6.3 | 6.3 | 7.5 | 9.6 |
| $^{56}Fe(n,p)$ | 1.6 | 1.4 | 2.1 | 1.1 | 1.4 |
| $^{197}Au(n,2n)$ | 6.7 | 6.3 | 6.3 | 8.5 | -0.3 |
| $^{24}Mg(n,p)$ | 3.9 | 6.1 | 6.1 | 6.1 | 8.3 |
| $^{58}Ni(n,x)^{57}Co$ | 2.6 | 2.8 | 4.6 | 16.9 | 1.4 |
| $^{90}Zr(n,2n)$ | -7.5 | -7.1 | -6.4 | -3.8 | -3.8 |
| $^{59}Co(n,\alpha)$ | 4.5 | 1.8 | -0.1 | 7.7 | 0.5 |
| $^{59}Co(n,2n)$ | 3.0 | 2.2 | 3.8 | 0.9 | 9.0 |
| $^{59}Co(n,p)$ | -4.8 | -4.5 | -5.0 | 0.0 | 0.8 |
| $^{51}V(n,\alpha)$ | 0.4 | - | - | - | - |
| $^{55}Mn(n,2n)$ | 7.1 | -0.1 | 10.9 | 6.0 | 6.0 |
| $^{169}Tm(n,2n)$ | -3.0 | -5.8 | -4.8 | -6.3 | - |
| $^{46}Ti(n,p)$ | -5.0 | 7.1 | -5.5 | -5.2 | -8.1 |

## 5 Conclusions

A large set of reaction rates was measured in the low-emission neutron generator. The $^{115}$In(n,n') is a suitable reaction for the determination of spectra contamination using low energy peaks. Neutron flux derived from indium dosimeter activation was shown to be in good agreement with the stilbene measurement.

It was also shown that the low-emission generator is usable for the measurement of the cross sections. The low neutron flux can be effectively compensated by the use of large foils with a large number of target nuclei. Different thicknesses of the foils can be easily treated by flux correction. With this approach, reaction rates are normalized to monitor flux and can be rescaled to a cross section. The $^{27}$Al(n,α) reaction can be recommended as the monitor. The obtained values of the experimental cross sections are in very good agreement with IRDFF-II dosimetric cross sections which can be understood as reference ones.

## Acknowledgments

The presented work has been realized within Institutional Support by the Ministry of Industry and Trade and with the use of the infrastructure Reactors LVR-15 and LR-0, which is financially supported by the Ministry of Education, Youth and Sports – project LM2015074, the SANDA project funded under H2020-EURATOM-1.1 contract 847552, and SP 2022/25 and SP 2023/036 projects supported by the Ministry of Education, Youth and Sports of the Czech Republic. Computational resources were supplied by the project "e-Infrastruktura CZ" (e-INFRA CZ LM2018140) supported by the Ministry of Education, Youth and Sports of the Czech Republic.

## References

[1] IAEA, Manual for troubleshooting and upgrading of neutron generators, IAEA-TECDOC-913, Vienna, 1996, ISSN: pp.1011-4289.

[2] Qian Li, Liyang Jiang, Chunlei Zhang, Xichao Ruan, Zhigang Ge, Measurement of the $^{59}$Co(n,2n)$^{58}$Co reaction cross section induced by 14.8 MeV neutrons, Applied Radiation and Isotopes, Vol.186, 2022, p. 110260,

[3] Jianfeng Liang, Feng Xie, Jie Bao, Xuesong Li, Quanlin Shi, Jianbo Shang, Xichao Ruan, Wengang Jiang, Gongshuo Yu, Xiongjun Chen, Tai Kang, Measurement of the 124Xe (n,2n) reaction cross section induced by 14.8 MeV neutron, Radiation Physics and Chemistry, Vol. 190, 2022

[4] L.W. Packer, M. Gilbert, S. Lilley, Integral Cross Section Measurements Around 14 MeV for Validation of Activation Libraries, Nuclear Data Sheets, Vol. 119, 2014, pp 173-175

[5] I.A Reyhancan, M Bostan, A Durusoy, a Elmalı, A Baykal, Y Özbir, Measurements of isomeric cross sections for (n,2n) reaction on 140Ce, 142Nd and 144Sm isotopes around 14 MeV, Annals of Nuclear Energy, Vol. 30, (2003), pp. 1539-1547

[6] F.M.D. Attar, G.T. Bholane, T.S. Ganesapandy, S.D. Dhole, V.N. Bhoraskar, Isomeric cross sections of the (n, α) reactions on the 90Zr, 93Nb and 92Mo isotopes measured for 13.73 MeV–14.77 MeV and estimated for 10 MeV–20 MeV neutron energies, Appl. Rad. And Isot., Vol. 184, 2022.

[7] Czarnecki S., Hardgrove C.J., Gasda P.J., Gabriel T.S.J., Starr M., Rice M.S., Frydenvang J., Wiens R.C., Rapin W., Nikiforov S., Lisov D., Litvak M., Calef F., Gengl H., Newsom H., Thompson L., Nowicki S., Identification and description of a silicic volcaniclastic layer in

Gale crater, Mars, using active neutron interrogation, J. Geophys. Res. Planets, 125 (3) (2020), Article e2019JE006180

[8] L.E. Heffern, C.J. Hardgrove, A. Parsons, E.B. Johnson, R. Starr, G. Stoddard, R.E. Blakeley, T. Prettyman, T.S.J. Gabriel, H. Barnaby, J. Christian, M.A. Unzueta, C. Tate, A. Martin, J. Moersch, Active neutron interrogation experiments and simulation verification using the SIngle-scintillator Neutron and Gamma-Ray spectrometer (SINGR) for geosciences, Nuclear Instruments and Methods in Physics Research Section A: Accelerators, Spectrometers, Detectors and Associated Equipment, Vol. 1020, (2021), p. 165883,

[9] Aleksandr Kavetskiy, Galina Yakubova, Stephen A. Prior, H. Allen Torbert, "Hot background" of the mobile inelastic neutron scattering system for soil carbon analysis, Appl. Rad. and Isot., Vol. 107, 2016, pp. 299-311

[10] A. Buffler, J. Tickner, Detecting contraband using neutrons: Challenges and future directions, Radiation Measurements, Volume 45, Issue 10, 2010, pp. 1186-1192,

[11] P. Andersson, T. Bjelkenstedt, E. Andersson Sundén, H. Sjöstrand, S. Jacobsson-Svärd, Neutron Tomography Using Mobile Neutron Generators for Assessment of Void Distributions in Thermal Hydraulic Test Loops, Physics Procedia, Volume 69, 2015, pp. 202-209,

[12] R. Adams, R. Zboray, M. Cortesi, H-.M. Prasser, Conceptual design and optimization of a plastic scintillator array for 2D tomography using a compact D–D fast neutron generator Appl. Rad. and Isot., Vol., 86 (2014), pp. 63-70

[13] Z. Ghani, S. Popovichev, P. Batistoni, L.W. Packer, A. Milocco, A. Cufar, D.J. Thomas, N.J. Roberts, L. Snoj, S. Jednorog, E. Laszynska et al., Characterisation of neutron generators and monitoring detectors for the in-vessel calibration of JET, Fusion Engineering and Design, Volume 136, Part A, 2018, pp 233-238

[14] Tomáš Czakoj, Michal Košťál, Marek Zmeškal, Evžen Novák, Filip Mravec, František Cvachovec, Jan Šimon, Martin Schulc, Radim Uhlář, Petr Alexa, Zdeněk Matěj, The characterization of D–T neutron generators in precise neutron experiments, Nuclear Instruments and Methods in Physics Research Section A: Accelerators, Spectrometers, Detectors and Associated Equipment, Vol. 1034, (2022), p. 166837,

[15] MP 320 neutron generator, 2021, https://www.thermofisher.com/order/catalog/product/1517021A?SID=srch-srp-1517021A#/1517021A?SID=srch-srp-1517021A. (Accessed 4 November 2022).

[16] G. Gandolfo, L. Lepore, Monte Carlo modeling of D-T neutron generator: improvement and experimental validation of a MCNPX input deck, in: High Perform. Comput. CRESCO Infrastruct. Res. Act. Results 2017, ENEA, 2018: p. 81. https://www.enea.it/en/publications/abstract/cresco-report-2017.

[17] D.A. Brown, M.B. Chadwick, R. Capote et al, "ENDF/B-VIII.0.0: The 8th Major Release of the Nuclear Reaction Data Library with CIELO-project Cross Sections, New Standards and Thermal Scattering Data", Nucl. Data Sheets, 148 (2018), pp. 1–142.

[18] J. F. Ziegler, M. D. Ziegler, and J. P. Biersack, "SRIM—The stopping and range of ions in matter", Nucl. Instrum. Methods Phys. Res. B, Beam Interact. Mater. At., Vol. 268, Nos. 11–12 (2010), pp. 1818–1823.

[19] H. Liskien and A. Paulsen, "Neutron production cross sections and energies for the reactions T(p,n)3He, D(d,n)3He, and T(d,n)4He", Atom. Data Nucl. Data, Vol. 11, No. 7 (1973), pp. 569–619.

[20] R. Uhlář, P. Alexa, O. Harkut, and P. Haroková, “Modified Texas Convention Method for Fast Neutron Flux Measurements“, IEEE Transaction on Nuclear Science, Vol. 67, No. 1 (2020), pp. 382-388.

[21] Cvachovec J., Cvachovec F.: Maximum Likelihood Estimation of a Neutron Spectrum and Associated Uncertainties, Advances in Military Technology, Vol.1, No. 2 January 2007, pp. 5 – 28

[22] Matěj Z., Košťál M., Majerle M., Ansorge M., Losa E., Zmeškal M., Schulc M., Šimon J., Štefánik M., Novák J., Koliadko D., Cvachovec F., Mravec F., Přenosil V., Zach V., Czakoj T., Rypar V., Capote R., The methodology for validation of cross sections in quasi monoenergetic neutron field, Nuclear Instruments and Methods in Physics Research, Section A: Accelerators, Spectrometers, Detectors and Associated Equipment, 1040, (2022), p. 167075

[23] Košťál M.; Rypar V.; Schulc M.; Losa E.; Baroň P.; Mareček M.; Uhlíř J.; Measurement of 75As(n,2n) cross section in well-defined spectrum of LR-0 special core, Ann. of. Nucl. En., Vol. 100, (2017), pp. 42 - 491

[24] E. Tomarchio, S.Rizzo, Coincidence-summing correction equations in gamma-ray spectrometry with p-type HPGe detectors, Radiation Physics and Chemistry, Vol. 80 (2011), pp. 318–323

[25] Trkov, A., Griffin, P.J., Simakov, S. et al., 2019. IRDFF-II: a New Neutron Metrology Library. Nucl. Data Sheets 163, pp. 1–108.

[26] O. Harkut, P. Alexa, and R. Uhlář, „Radiocaesium Contamination of Mushrooms at High- and Low-Level Chernobyl Exposure Sites and Its Consequences for Public Health“, Life, Vol. 11 (2021), 1370, pp. 1-9, https://doi.org/10.3390/life11121370

[27] S. Agostinelli, J. Allison, K. Amako, J. Apostolakis, H. Araujo, P. Arce, M. Asai, D. Axen, S. Banerjee, G. Barrand, F. Behner, L. Bellagamba, J. Boudreau, L. Broglia, A. Brunengo, H. Burkhardt, S. Chauvie, J. Chuma, R. Chytracek, G. Cooperman, G. Cosmo, P. Degtyarenko, A. Dell’Acqua, G. Depaola, D. Dietrich, R. Enami, A. Feliciello, C. Ferguson, H. Fesefeldt, G. Folger, F. Foppiano, A. Forti, S. Garelli, S. Giani, R. Giannitrapani, D. Gibin, J.J. Gómez Cadenas, I. González, G. Gracia Abril, G. Greeniaus, W. Greiner, V. Grichine, A. Grossheim, S. Guatelli, P. Gumplinger, R. Hamatsu, K. Hashimoto, H. Hasui, A. Heikkinen, A. Howard, V. Ivanchenko, A.S. Johnson, F.W. Jones, J. Kallenbach, N. Kanaya, M. Kawabata, Y. Kawabata, M. Kawaguti, S. Kelner, P. Kent, A. Kimura, T. Kodama, R. Kokoulin, M. Kossov, H. Kurashige, E. Lamanna, T. Lampén, V. Lara, V. Lefebure, F. Lei, M. Liendl, W. Lockman, F. Longo, S. Magni, M. Maire, E. Medernach, K. Minamimoto, P. Mora de Freitas, Y. Morita, K. Murakami, M. Nagamatu, R. Nartallo, P. Nieminen, T. Nishimura, K. Ohtsubo, M. Okamura, S. O’Neale, Y. Oohata, K. Paech, J. Perl, A. Pfeiffer, M.G. Pia, F. Ranjard, A. Rybin, S. Sadilov, E. Di Salvo, G. Santin, T. Sasaki, N. Savvas, Y. Sawada, S. Scherer, S. Sei, V. Sirotenko, D.A. Smith, N. Starkov, H. Stoecker, J. Sulkimo, M. Takahata, S. Tanaka, E. Tcherniaev, E. Safai Tehrani, M. Tropeano, P. Truscott, H. Uno, L. Urban, P. Urban, M. Verderi, A. Walkden, W. Wander, H. Weber, J.P. Wellisch, T. Wenaus, D.C. Williams, D. Wright, T. Yamada, H. Yoshida, D. Zschiesche, Geant4—a simulation toolkit, Nucl. Instruments Methods Phys. Res. Sect. a Accel. Spectrometers, Detect. Assoc. Equip. 506 (2003), pp. 250–303. https://doi.org/10.1016/S0168-9002(03)01368-8.

[28] E. Mendoza, D. Cano-Ott, T. Koi, C. Guerrero, New Standard Evaluated Neutron Cross Section Libraries for the GEANT4 Code and First Verification, IEEE Trans. Nucl. Sci. 61 (2014), pp. 2357–2364. https://doi.org/10.1109/TNS.2014.2335538.

[29] E. Mendoza, D. Cano-Ott, C. Guerrero, R. Capote, New evaluated neutron cross section libraries for the GEANT4 code, Vienna, 2012. https://www-nds.iaea.org/publications/indc/indc-nds-0612.pdf (accessed March 7, 2022).

[30] E. Mendoza, D. Cano-Ott, Update of the Evaluated Neutron Cross Section Libraries for the Geant4 Code, Vienna, 2018. https://www-nds.iaea.org/publications/indc/indc-nds-0758.pdf (accessed July 1, 2021).

[31] Košťál M., Losa E., Matěj Z., Juříček V., Harutyunyan D., Huml O., Štefánik M., Cvachovec F., Mravec F., Schulc M., Czakoj T., Rypar V.; Characterization of mixed N/G beam of the VR-1 reactor, Annals of Nuclear Energy, 122, (2018), pp. 69 - 78

[32] Kostal M., Schulc M.; Burianova N.; Harutyunyan, D.; Losa E.; Rypar V.; Measurement of various monitors reaction rate in a special core at LR-0 reactor, Ann. of Nucl. En., 112, pp. 759-768, (2018)

[33] Filatenkov A. A.; Neutron activation cross sections measured at KRI in neutron energy region 13.4 - 14.9 MeV, Rept: USSR report to the I.N.D.C., No.0460. (2016).

[34] Ikeda Y., Konno C., Oishi K., Nakamura T., Miyade H., Kawade K., Yamamoto H., Katoh T.; Activation cross section measurements for fusion reactor structural materials at neutron energy from 13.3 to 15.0 MeV using FNS facility, JAERI Reports, No.1312, (1988).

[35] Zhang Y., Zhao L., Kong X., Liu R., Jiang L.; Cross sections for (n,2n) and (n,alpha) reactions on 55Mn isotope around neutron energy of 14 MeV, Radiation Physics and Chemistry, Vol.81, p.1563. (2012).

[36] Konno C., Ikeda Y., Oishi K., Kawade K., Yamamoto H., Maekawa H.; Activation Cross section measurements at neutron energy from 13.3 to 14.9 MeV, JAERI Reports, No.1329. (1993).

[37] Ghorai S.K., Hudson C.G., Alford W.L.; The Excitation Function for the 89Y(n,2n)88Y Reaction, Nuclear Physics, Section A, Vol.266, p.53. (1976). DOI: 10.1016/0375-9474(76)90281-5. NSR: 1976GH02.

[38] Molla N.I., Basunia S., Miah M.R., Hossain S.M., Rahman M.M., Spellerberg S., Qaim S.M.; Radiochemical Study of 45Sc(n,p)45Ca and 89Y(n,p)89Sr Reactions in the Neutron Energy Range of 13.9 to 14.7 MeV, Radiochimica Acta, Vol.80, p.189. (1998). DOI: 10.1524/ract.1998.80.4.189. NSR: 1998MO15.

[39] Iguchi T., Nakata K,, Nakazawa M.; Improvement of Accuracy of Zr/Nb Activation-Rate Ratio Method for D-T Neutron Source Energy Determination, Journal of Nuclear Science and Technology, 24:12, 1076-1079. (1987). DOI: 10.1080/18811248.1987.9733544.

[40] Prestwood R.J., Bayhurst B.P.; (n, 2n) Excitation Functions of Several Nuclei from 12.0 to 19.8 MeV, Physical Review, vol. 121, Issue 5, pp. 1438-1441. (1961).

[41] Chiadli A., Paic G.; Cross-section of (n,2n) reaction on 93Nb and 90Zr, Univ. Mohammed V, Rabat, Annual Report, No.5, p.13. (1982).

[42] Nethaway D.R.; Cross sections for several (n, 2n) reactions at 14 Mev. Nuclear Physics A, Volume 190, Issue 3, Pages 635-644. (1972). ISSN 0375-9474.

[43] Csikai J.; Study of excitation functions around 14 MeV neutron energy, Conf.on Nucl.Data for Sci.and Technol.,Antwerp 1982, p.414. (1982).

[44] Zhu Ch., Chen Y.,Mou Y., Zheng P., He T., Wang X., An L., Guo H.; Measurements of (n, 2n) Reaction Cross Sections at 14 MeV for Several Nuclei, Nuclear Science and Engineering, 169:2, 188-197. (2011). DOI: 10.13182/NSE10-35.